\documentclass[10pt,twocolumn,twoside]{IEEEtran}
\usepackage{cite}
\usepackage{threeparttable}
\usepackage{graphicx}
\usepackage{picinpar}
\usepackage{color}
\usepackage[cmex10]{amsmath}
\usepackage{amsmath,amsfonts,amssymb}
\usepackage{subfigure}
\usepackage{algorithm}
\usepackage{algorithmic}
\usepackage{stfloats}
\usepackage{bm}
\usepackage{amsthm}

\begin{document}

\title{Super-Resolution Channel Estimation for MmWave Massive MIMO with Hybrid Precoding}

\author{
\IEEEauthorblockN{Chen Hu, Linglong Dai,~{\it Senior Member,~IEEE},~Talha Mir,~Zhen Gao,~{\it Member,~IEEE},\\ and Jun Fang,~{\it Senior Member,~IEEE}}

\thanks{Copyright {\textcopyright} 2015 IEEE. Personal use of this material is permitted. However, permission to use this material for any other purposes must be obtained from the IEEE by sending a request to pubs-permissions@ieee.org.}
\thanks{C. Hu, L. Dai and T. Mir are with Department of Electronic Engineering, Tsinghua University, Beijing 100084, P. R. China (E-mail: huc16@mails.tsinghua.edu.cn, daill@tsinghua.edu.cn, bah15@mails.tsinghua.edu.cn).}
\thanks{Z. Gao is with Advanced Research Institute for Multidisciplinary Science (ARIMS), Beijing Institute of Technology, Beijing 100081, P. R. China (E-mail: gaozhen16@bit.edu.cn).}
\thanks{J. Fang is with the National Key Laboratory of Science and Technology on Communications, University of Electronic Science and Technology of China, Chengdu 611731, P. R. China (E-mail: JunFang@uestc.edu.cn).}
\vspace{-8mm}

\thanks{This work was supported by the National Natural Science Foundation of China for Outstanding Young Scholars (Grant No. 61722109), the National Natural Science Foundation of China (Grant No. 61571270), and the Royal
Academy of Engineering through the UK-China Industry Academia Partnership Programme Scheme (Grant No. UK-CIAPP$\backslash$49).}
}

\maketitle
\begin{abstract}
 Channel estimation is challenging for millimeter-wave (mmWave) massive MIMO with hybrid precoding, since the number of radio frequency (RF) chains is much smaller than that of antennas. Conventional compressive sensing based channel estimation schemes suffer from severe resolution loss due to the channel angle quantization. To improve the channel estimation accuracy, we propose an iterative reweight (IR)-based super-resolution channel estimation scheme in this paper. By optimizing an objective function through the gradient descent method, the proposed scheme can iteratively move the estimated angle of arrivals/departures (AoAs/AoDs) towards the optimal solutions, and finally realize the super-resolution channel estimation. In the optimization, a weight parameter is used to control the tradeoff between the sparsity and the data fitting error. In addition, a singular value decomposition (SVD)-based preconditioning is developed to reduce the computational complexity of the proposed scheme. Simulation results verify the better performance of the proposed scheme than conventional solutions.
\end{abstract}

\IEEEpeerreviewmaketitle
\vspace{-2.5mm}
\section{Introduction}\label{S1}
Millimeter-wave (mmWave) massive MIMO has been recognized as a promising technology for future 5G wireless communications\cite{G1}. To reduce the hardware cost and power consumption, hybrid precoding has been proposed for practical mmWave massive MIMO systems, where hundreds of antennas are driven by a much smaller number of radio frequency (RF) chains\cite{G2,G3}. The analog and digital co-design in hybrid precoding requires accurate channel state information. However, the digital baseband cannot directly access all antennas due to the small number of RF chains, so it is difficult to accurately estimate the high-dimensional MIMO channel \cite{FGao,G6}.

Several novel channel estimation schemes have been recently proposed for mmWave massive MIMO with hybrid precoding\cite{G6,G7,ABP,G8,G10}. Specifically, \cite{G6,G7} proposed the adaptive codebook-based channel sounding scheme, where the transmitter and receiver search for the best beam pair by adjusting the predefined precoding and combining codebooks. However, the channel estimation resolution is limited by the codebook size. \cite{ABP} was able to achieve better angle estimation by performing an amplitude comparison with respect to the auxiliary beam pair. On the other hand, by exploiting the angular channel sparsity, the on-grid compressive sensing based methods \cite{G8,G10} could estimate the channel with reduced training overhead. However, such solutions assumed that the angle of arrivals/departures (AoAs/AoDs) lie in discrete points in the angle domain (i.e., ``on-grid" AoAs/AoDs), while the actual AoAs/AoDs are continuously distributed (i.e., ``off-grid" AoAs/AoDs) in practice. The assumption of on-grid AoAs/AoDs results in the power leakage problem, which severely degrades the channel estimation accuracy. To solve this resolution limitation caused by the on-grid angle estimation, we propose an iterative reweight (IR)-based super-resolution channel estimation scheme to estimate the off-grid AoAs/AoDs.\footnotemark[1]

\footnotetext[1]{
Simulation codes are provided to reproduce the results presented in this paper: http://oa.ee.tsinghua.edu.cn/dailinglong/publications/publications.html.
}

Specifically, we iteratively optimize the estimates of AoAs/AoDs, to decrease the weighted summation of the sparsity and the data fitting error. The weight controlling the tradeoff between the sparsity and the data fitting error, is iteratively updated to avoid over-fitting or under-fitting.
{Since the estimated AoAs/AoDs can be moved from the initial angle-domain grids towards the actual off-grid AoAs/AoDs, the proposed scheme is able to achieve the super-resolution channel estimation.}
In addition, we propose a singular value decomposition (SVD)-based preconditioning method to reduce the computational complexity of the proposed scheme, which is realized by reducing the number of initial candidates of AoAs/AoDs in the IR procedure. Simulation results show that the proposed IR-based super-resolution channel estimation can achieve better performance than conventional solutions.

The contributions of this paper are the follows. We propose a novel IR-based super-resolution channel estimation scheme for mmWave massive MIMO with hybrid precoding. Comparing with the state-of-art schemes schemes such as those in \cite{G7,ABP,G8}, we can achieve super-resolution channel estimation, which means substantially improved estimation accuracy. Moreover, the proposed SVD based preconditioning significantly reduces the computational complexity of the IR procedure, and makes the method practical in mmWave channel estimation.

\emph{Notation: }In this paper, the boldface lower and upper-case symbols denote vectors and matrices. $(\cdot )^{T}$, $(\cdot )^{H}$ and $(\cdot )^{-1}$ denote the transpose, the conjugate transpose, and the inverse of a matrix, respectively. ${\rm{diag}}({\bf{x}})$ is the diagonal matrix with the vector ${\bf{x}}$ on its diagonal. The $\ell_{0}$-norm, $\ell_{2}$-norm, and Frobenius norm are given by $\|\cdot\|_0$, $\|\cdot\|_2$, and $\|\cdot\|_{\rm F}$, respectively.

\vspace{-1.5mm}
\section{System Model}\label{S2}
{We consider a hybrid-precoding mmWave massive MIMO with arbitrary array geometry.} Let $N_{\rm T}$, $N_{\rm R}$, $N_{\rm T}^{\rm RF}$, and $N_{\rm R}^{\rm RF}$ be the number of transmit antennas, receive antennas, transmitter RF chains, and receiver RF chains, respectively. For practical mmWave massive MIMO with hybrid precoding, the number of RF chains is much smaller than that of antennas, i.e., $N_{\rm T}^{\rm RF} < N_{\rm T}$, $N_{\rm R}^{\rm RF} < N_{\rm R}$ \cite{G1,G2,G3}. The system model can be given by
\begin{equation}\label{equ:y}
\vspace*{-1mm}
{\bf{r}} = {\bf{Q}}^{H} {\bf{HPs}} + {\bf{n}},
\end{equation}
where ${\bf{r}} \in {\mathbb{C}^{N_{\rm R}^{\rm RF} \times 1}}$ is the received signal, ${\bf{Q}} \in {\mathbb{C}^{N_{\rm R} \times N_{\rm R}^{\rm RF}}}$ is the hybrid combining matrix, ${\bf{H}} \in {\mathbb{C}^{N_{\rm R} \times N_{\rm T}}}$ is the channel matrix, ${\bf{P}} \in {\mathbb{C}^{N_{\rm T} \times N_{\rm T}^{\rm RF}}}$ is the hybrid precoding matrix, ${\bf{s}} \in {\mathbb{C}^{N_{\rm T}^{\rm RF} \times 1}}$ is the transmitted signal, and ${\bf{n}} \in {\mathbb{C}^{N_{\rm R}^{\rm RF} \times 1}}$ is the received noise after combining.

{
The channel model
\vspace*{-1.5mm}
\begin{equation}\label{equ:H}
{\bf{H}} = \sum\limits_{l = 1}^L {{z_l}{{\bf{a}}_{\rm R}}\!\left(\phi^{\rm azi} _{{\rm R},l},\phi^{\rm ele} _{{\rm R},l}\right) {\bf{a}}_{\rm T}^{H}\!\left(\phi^{\rm azi} _{{\rm T},l},\phi^{\rm ele} _{{\rm T},l}\right)}
\vspace*{-1.5mm}
\end{equation}
is widely adopted in mmWave massive MIMO systems, and it is regarded almost unchanged within the channel coherence time for channel estimation \cite{G7,ABP,G8}, where $L$ is the number of propagation paths, $L \ll {\rm min}\left(N_{\rm R},N_{\rm T}\right)$, $z_l$, $\phi^{\rm azi} _{{\rm R},l}~(\phi^{\rm ele} _{{\rm R},l})$ and $\phi^{\rm azi} _{{\rm T},l}~(\phi^{\rm ele} _{{\rm T},l})$ are the complex path gain, the azimuth (elevation) AoA and AoD of the $l$-th path, respectively. ${{\bf{a}}_{\rm R}}\!\left(\phi^{\rm azi} _{{\rm R},l},\phi^{\rm ele} _{{\rm R},l}\right)$ and ${{\bf{a}}_{\rm T}}\!\left(\phi^{\rm azi} _{{\rm T},l},\phi^{\rm ele} _{{\rm T},l}\right)$ are the steering vector at the receiver and the steering vector at the transmitter, respectively. These steering vectors depend on the array geometry. Ignoring the subscripts without loss of generality, for the typical $N_1\times N_2$ uniform planar arrays (UPAs), ${{\bf{a}}}\!\left(\phi^{\rm azi} _{l},\phi^{\rm ele} _{l} \right)$ is given by \cite{G3}
\vspace*{-2mm}
\begin{equation}
\begin{aligned}
&{\!{{\bf{a}}}\!\left(\phi^{\!\rm azi}\!\!,\!\phi^{\!\rm ele} \right) \!\!=\!\! {\left[1,{{e^{j2\pi d\sin\!{\phi^{\!\rm azi}}\!\!\sin\!{\phi^{\!\rm ele}}\!\!/\!\lambda }}},{\cdots},{{e^{j2\pi   \left(\!{N_{\!1} \!-\! 1}\! \right)d\sin\! {\phi^{\!\rm azi}}\!\!\sin\!{\phi^{\!\rm ele}}\!\!/\!\lambda }}}\! \right]^{\!{T}}}\!\!}\\
&~~~~~~~~~~~~~{\otimes}{\left[1,{{e^{j2\pi d\cos{\phi^{\rm ele}}\!/\!\lambda }}},{\cdots},{{e^{j2\pi   \left({N_{\!2} - 1} \right)d\cos{\phi^{\rm ele}}\!/\!\lambda }}} \right]^{\!{T}}},
\end{aligned}
\vspace*{-1.5mm}
\end{equation}
where $d$ is the antenna spacing, $\lambda$ is the wavelength, $\otimes$ denotes the Kronecker product. For uniform linear arrays (ULAs), the steering vector is only determined by one angle \cite{G7}
\vspace*{-1.5mm}
\begin{equation}
\vspace*{-1mm}
{{{\bf{a}}}\left(\phi\right) = {\left[1,{{e^{j2\pi d\sin{\phi}/\lambda }}},{\cdots},{{e^{j2\pi \left({N- 1} \right)d\sin{\phi}/\lambda }}}\right]^{\!{T}}}}.
\vspace*{-0.5mm}
\end{equation}

By defining the normalized spacial angles by ${\theta^{\rm azi}}\buildrel \Delta \over = d\sin {\phi^{\rm azi}}\sin {\phi^{\rm ele}}/\lambda$ and ${\theta^{\rm ele}}\buildrel \Delta \over = d\cos {\phi^{\rm ele}}/\lambda$, the mmWave channel matrix $\bf H$ in (\ref{equ:H}) can be also written as
\vspace*{-1.5mm}
\begin{equation}\label{equ:H1}
{\bf{H}} = {{\bf{A}}_{\rm R}}\!\left( {{\bm{\theta} _{\rm R}}} \right){\rm{diag}}\left( {\bf{z}} \right){\bf{A}}_{\rm T}^{H}\!\left( {{\bm{\theta} _{\rm T}}} \right),
\vspace*{-1mm}
\end{equation}
where ${\bf z}\!=\!\left[z_1,z_2,\cdots,z_L\right] ^T$, ${{\bm{\theta} _{\rm R}}} \!=\! \left[{\theta^{\rm azi}_{{\rm R}\!,1}},{\theta^{\rm ele}_{{\rm R}\!,1}},{\theta^{\rm azi}_{{\rm R}\!,2}},{\theta^{\rm ele}_{{\rm R}\!,2}},{\cdots},\right.$\\$\left.{\theta^{\rm azi}_{{\rm R}\!,L}},{\theta^{\rm ele}_{{\rm R}\!,L}}\right]^{T}$, ${{\bm{\theta} _{\rm T}}} \!= \! \left[{\theta^{\rm azi}_{{\rm T}\!,1}},{\theta^{\rm ele}_{{\rm T}\!,1}},{\theta^{\rm azi}_{{\rm T}\!,2}},{\theta^{\rm ele}_{{\rm T}\!,2}},{\cdots},{\theta^{\rm azi}_{{\rm T}\!,L}},{\theta^{\rm ele}_{{\rm T}\!,L}}\right]^{T}$, ${{\bf{A}}_{\!\rm R}}\!\left( {{\bm{\theta} _{\rm R}}} \right)\! =\! \left[{{{\bf{a}}_{\rm R}}({\theta^{\rm azi}_{{\rm R},1}},{\theta^{\rm ele}_{{\rm R},1}})}~{{{\bf{a}}_{\rm R}}({\theta^{\rm azi}_{{\rm R},2}},{\theta^{\rm ele}_{{\rm R},2}})}~{\cdots}~{{{\bf{a}}_{\rm R}}({\theta^{\rm azi}_{{\rm R},L}},{\theta^{\rm ele}_{{\rm R},L}})} \right]$, ${{\bf{A}}_{\!\rm T}}\!\left( {{\bm{\theta} _{\rm T}}} \right)\! =\!\left[{{{\bf{a}}_{\rm T}}({\theta^{\rm azi}_{{\rm T},1}},{\theta^{\rm ele}_{{\rm T},1}})}~{{{\bf{a}}_{\rm T}}({\theta^{\rm azi}_{{\rm T},2}},{\theta^{\rm ele}_{{\rm T},2}})}~{\cdots}~{{{\bf{a}}_{\rm T}}({\theta^{\rm azi}_{{\rm T},L}},{\theta^{\rm ele}_{{\rm T},L}})} \right]$.}

Denote ${\bf{x}} = {\bf{Ps}} \in {\mathbb{C}^{N_{\rm T} \times 1}}$, where the $i$-th element of $\bf{x}$ is the transmitted signal at the $i$-th transmit antenna. Suppose that the transmitter sends $N_{\rm X}~(N_{\rm X}<N_{\rm T})$ different pilot sequences ${\bf{x}}_1, {\bf{x}}_2, \cdots, {\bf{x}}_{N_{\rm X}}$. Since the number of RF chains is smaller than the required dimension of received pilot sequence, for each transmit pilot sequence ${\bf{x}}_p~(1 \!\le\! p\! \le\! N_{\rm X})$, we use $M$ time slots to obtain an $N_{\rm Y}$-dimensional received pilot sequence ${\bf{y}}_p$, where $N_{\rm Y}=MN_{\rm R}^{\rm RF}$. Thus, the training overhead is $T=MN_{\rm X}$. In the $m$-th time slot, we use the combining matrix ${\bf W}_m$ to obtain an $N_{\rm R}^{\rm RF}$-dimensional received pilot sequence
\vspace*{-1mm}
\begin{equation}
{\bf y}_{p,m}={\bf W}_m^{H}{\bf H}{\bf x}_p+{\bf n}_{p,m}.
\vspace*{-0.5mm}
\end{equation}
By collecting the received pilots in the $M$ time slots, we have ${{\bf y}_p} = {\bf W}^{H}{\bf Hx}_p + {\bf n}_p$,
where ${{\bf y}_p}=[{{\bf y}_{p,1}^{T}},~{{\bf y}_{p,2}^{T}},~\cdots,~{{\bf y}_{p,M}^{T}}]^{T}\in {\mathbb{C}^{N_{\rm Y}\times 1}}$, ${{\bf W}}=[{{\bf W}_{1}},{{\bf W}_{2}},\cdots,{{\bf W}_{M}}]\in {\mathbb{C}^{N_{\rm R}\times N_{\rm Y}}}$, ${\bf n}_p\in {\mathbb{C}^{N_{\rm Y} \times 1}}$ is the noise. By defining ${\bf{Y}}=[{\bf{y}}_1,{\bf{y}}_2, \cdots, {\bf{y}}_{N_{\rm X}}]$, ${{\bf{X}}=\left[{\bf{x}}_1, {\bf{x}}_2, \cdots, {\bf{x}}_{N_{\rm X}} \right]}$, ${{\bf{N}}=\left[{\bf{n}}_1, {\bf{n}}_2, \cdots, {\bf{n}}_{N_{\rm X}} \right]}$, we have
\vspace*{-0.5mm}
\begin{equation}\label{equ:Y}
{\bf Y} = {\bf W}^{H}{\bf HX + N}.
\vspace*{-0.5mm}
\end{equation}

The estimation of the channel matrix $\bf H$ in (\ref{equ:Y}) is equivalent to the estimation of the number of paths, the normalized spacial angles $({{\bm\theta} _{{\rm T}}}$, ${{\bm\theta} _{{\rm R}}})$, and path gains ${\bf z}$ for all $L$ paths. Due to the angle-domain sparsity of the channel matrix $\bf H$, the sparse channel estimation problem can be formulated as
\vspace*{-0.5mm}
\begin{equation}\label{equ:l0opt}
\begin{aligned}
\mathop {\min }\limits_{{\bf{\hat z}},{\bm{\hat \theta} _{\rm R}},{\bm{\hat \theta} _{\rm T}}}   {\left\| {\bf{\hat z}} \right\|_0}, ~~{\rm{s}}{\rm{.t}}{\rm{.}} ~ {\left\| {{\bf{Y}} - {\bf W}^H{\bf {\hat H}X}} \right\|_{\rm F}} \le \varepsilon,
\end{aligned}
\end{equation}
where ${\left\| {\bf{\hat z}} \right\|_0}$ is the number of non-zero elements of $\bf \hat z$, which means the estimated number of paths $\hat L$, ${\bf {\hat H}}$ is the estimated channel matrix, and $\varepsilon$ is the error tolerance parameter \cite{G10}.

\section{Proposed IR-Based Super-Resolution Channel Estimation}\label{S3}

\subsection{Proposed Optimization Formulation}\label{IRFormulation}

The main difficulty in solving (\ref{equ:l0opt}) lies in the fact that the $l_0$-norm is not computationally efficient for finding the optimal solution. By replacing the $l_0$-norm with a log-sum function \cite{G12}, we have
\vspace*{-0.5mm}
\begin{equation}\label{equ:logsumopt}
\begin{aligned}
\mathop {\min }\limits_{{\bf{z}},{{\bm{\theta}} _{\rm R}},{{\bm{\theta}} _{\rm T}}}   F\!\left( {\bf{z}} \right) \buildrel \Delta \over = \sum\limits_{l = 0}^L {\log \left( {{{\left| {{z_l}} \right|}^2} + \delta } \right)},~{\rm{s}}{\rm{.t}}{\rm{.}} {\left\| {{\bf{Y}}\!-\!{\bf W\!}^H\!{\bf {\hat H}X}} \right\|_{\rm F}}\! \le \varepsilon,
\end{aligned}
\end{equation}
where $\delta >0$ ensures that the logarithmic function is well-defined \cite{G12}, ${\bf \hat H}$ is determined by the parameters ${\bf{z}},{{\bm{\theta}} _{\rm R}}$ and ${{\bm{\theta}} _{\rm T}}$ defined in (\ref{equ:H1}). By adding a regularization parameter $\lambda >0$, we can further formulate the problem (\ref{equ:logsumopt}) as a unconstrained optimization problem:
\vspace*{-0.5mm}
\begin{equation}\label{equ:G}
\mathop {\min }\limits_{{\bf{z}},{{\bm{\theta}} _{\rm R}},{{\bm{\theta}} _{\rm T}}}  G\left( {{\bf{z}},{{\bm{\theta}} _{\rm R}},{{\bm{\theta}} _{\rm T}}} \right) \buildrel \Delta \over = \sum\limits_{l = 1}^L {\log \!\left( {{{\left| {{z_l}} \right|}^2} \!+\! \delta } \right)}  + \lambda {\left\| {{\bf{Y}} \!-\! {\bf W}^{\!H}\!{\bf {\hat H}X}} \right\|^2_{\rm F}}.
\end{equation}
Moreover, by using an iterative surrogate function instead of the log-sum function, the minimization of $G\left( {{\bf{z}},{{\bm{\theta}} _{\rm R}},{{\bm{\theta}} _{\rm T}}} \right)$ is equivalent to the minimization of the surrogate function\cite{G12}:
\begin{equation}\label{equ:iteropt}
\mathop {\min }\limits_{{\bf{z}},{{\bm{\theta}} _{\rm R}},{{\bm{\theta}} _{\rm T}}}  {S^{(i)}}\!\!\left( {{\bf{z}},{{\bm{\theta}} _{\rm R}},{{\bm{\theta}} _{\rm T}}} \right) \buildrel \Delta \over = \lambda ^{-1}{{\bf{z}}^{\rm{H}}}{{\bf{D}}^{(i)}}{\bf{z}} + {\left\| {{\bf{Y}} - {\bf W}^H{\bf {\hat H}X}} \right\|^2_{\rm F}},
\end{equation}
where ${{\bf{D}}^{(i)}}$ is defined as
\begin{equation}
{{\bf{D}}^{(i)}} \buildrel \Delta \over = {\rm{diag}}\left(
   {\frac{1}{{{{\left| {\hat z_1^{(i)}} \right|}^2} \!\!+\! \delta }}}~~{\frac{1}{{{{\left| {\hat z_2^{(i)}} \right|}^2} \!\!+\! \delta }}}~{\cdots}~ {\frac{1}{{{{\left| {\hat z_L^{(i)}} \right|}^2} \!\!+\! \delta }}}\right),
\end{equation}
where ${\hat {\bf{z}}^{(i)}}$ is the estimate of ${\bf{z}}$ at the $i$-th iteration. 

Then, as proved in Appendix \ref{AppendixA}, we can optimize (\ref{equ:iteropt}) with regard to the path gains $\bf{z}$, to find the optimal point of $\bf \hat z$ and the corresponding optimal value of $S^{(i)}$ as follows:
\vspace{-1mm}
\begin{equation}\label{equ:zopt}
\begin{aligned}
{{\bf{z}}_{\rm opt}^{(i)}}\!\left( {{{\bm{\theta}} _{\rm R}},{{\bm{\theta}} _{\rm T}}} \right) \buildrel \Delta \over = &\arg\mathop{\min }\limits_{{\bf{z}}} ~S^{(i)}\!\left({\bf{z}}, {{{\bm{\theta}} _{\rm R}},{{\bm{\theta}} _{\rm T}}} \right)\\
\!=& {\left( {{\lambda ^{ \!- 1}}{{\bf{D}}^{(i)}}\! +\! \sum\limits_{p = 1}^{N_{\rm X}} {{\bf{K}}_p^{{H}}{{\bf{K}}_p}} } \right)^{ \!\!\!- 1}}\!\!\left( {\sum\limits_{p = 1}^{N_{\rm X}} {{\bf{K}}_p^{{H}}{{\bf{y}}_p}} } \right),
\end{aligned}
\end{equation}
\vspace{-5mm}
\begin{equation}\label{equ:sopt}
\begin{aligned}
    &{{S}_{\rm opt}^{(i)}}\!\left( {{{\bm{\theta}} _{\rm R}},{{\bm{\theta}} _{\rm T}}} \right) \buildrel \Delta \over = \mathop {\min }\limits_{{\bf{z}}} ~S^{(i)}\!\left({\bf{z}}, {{{\bm{\theta}} _{\rm R}},{{\bm{\theta}} _{\rm T}}} \right)\\=& \!-\! \!\left( {\sum\limits_{p = 1}^{N_{\rm X}} \! {{\bf{y}}_{\!p}^{\!{H}\!}{{\bf{K}}_{\!p}}} }\! \right) \!\!\cdot\!\! {\left( \!{{\lambda ^{\! - \!\! 1}}\!{{\bf{D}}^{\!(i)}}\!\! + \!\!\sum\limits_{p = 1}^{N_{\rm X}} \!{{\bf{K}}_{\!p}^{\!{H}\!}{{\bf{K}}_{\!p}}} } \!\!\right)^{ \!\!\!- \!1}}
    \!\!\!\!\cdot\!\!\left( {\sum\limits_{p = 1}^{N_{\rm X}} \!{{\bf{K}}_{\!p}^{\!{H}\!}{{\bf{y}}_{\!p}}} }\!\! \right)\! \!+\!\! \sum\limits_{p = 1}^{N_{\rm X}}\! {{\bf{y}}_{\!p}^{\!{H}\!}{{\bf{y}}_{\!p}}},
\end{aligned}
\end{equation}
where ${{\bf{K}}_{p}} = {\bf{W}}^H{{\bf{A}}_{\rm R}}{\rm{diag}}\left( {{\bf{A}}_{\rm T}^{{H}}{{\bf{x}}_p}} \right)$. After that, we only need to optimize the normalized spatial angles $ {{\bm{\theta}} _{\rm R}}$ and ${{\bm{\theta}} _{\rm T}}$ in (\ref{equ:sopt}), which will be discussed in the next subsection.

\subsection{IR-Based Super-Resolution Channel Estimation}\label{IRM}
In the previous subsection, we have already simplified the constrained optimization problem (\ref{equ:l0opt}) to an unconstrained angle optimization problem (\ref{equ:sopt}). To solve this reformulated problem, now we propose an IR-based super-resolution channel estimation scheme as described in {\bf Algorithm \ref{alg:Framwork}}.

\begin{algorithm}[!t]
{
\renewcommand{\algorithmicrequire}{\textbf{Input:}}
\renewcommand\algorithmicensure {\textbf{Output:} }
\caption{IR-based super-resolution channel estimation}
\label{alg:Framwork}
\begin{algorithmic}[1]
\small
\REQUIRE
 Noisy received signals ${\bf Y}$, transmit pilot signals ${\bf X}$, combining matrix ${\bf W}$, initial on-grid AoAs and AoDs ${\bm{\hat \theta}} _{\rm R}^{(0)}$, ${\bm{\hat \theta}} _{\rm T}^{(0)}$, pruning threshold $z_{\rm th}$ and termination threshold ${\bf {\varepsilon}}_{\rm th}$.

\ENSURE
 Estimated AoAs/AoDs and path gains of all paths.
\STATE {Initialize ${{\bf{\hat z}}^{(0)}} = {{\bf{z}}_{\rm opt}}\left( {{\bm{\hat \theta }}_{\rm R}^{(0)},{\bm{\hat \theta }}_{\rm T}^{(0)}} \right)$ according to (\ref{equ:zopt}).}

\REPEAT
\STATE {Update $\lambda$ by (\ref{equ:lambda}).}
\STATE {Construct the function ${{S}_{\rm opt}^{(i)}}\!\left( {{{\bm{\theta}} _{\rm R}},{{\bm{\theta}} _{\rm T}}} \right)$ by (\ref{equ:sopt}).}
\STATE {Search for new angle estimates ${\bm{\hat \theta}} _{\rm R}^{(i + 1)}$, ${\bm{\hat \theta}} _{\rm T}^{(i + 1)}$ by (\ref{equ:grad}).}
\STATE {Estimate the path gains ${{\bf{\hat z}}^{(i+1)}}$} according to (\ref{equ:zopt}).
\STATE {Prune path $l$ if ${{{\hat z}_{l}}^{(i+1)}} < z_{\rm th}$.}
\UNTIL{${L}^{(i)}={L}^{(i+1)}$ and ${\left\| {{{\bf{z}}^{(i+1)}} - {{\bf{z}}^{(i)}}} \right\|_2} < {\varepsilon _{\rm th}}.$}

\STATE {${{\bm{\hat \theta }}_{\rm R}} = {\bm{\hat \theta }}_{\rm R}^{(\rm last)},{{\bm{\hat \theta }}_{\rm T}} = {\bm{\hat \theta }}_{\rm T}^{(\rm last)},{\bf{\hat z}} = {{\bf{\hat z}}^{(\rm last)}}$.}
\end{algorithmic}
}
\end{algorithm}

The objective function ${{{S}}^{(i)}}\left({{\bf z}}, {{{\bm \theta} _{\rm R}},{{\bm \theta} _{\rm T}}} \right)$ is the weighted sum of two parts: ${{\bf{z}}^{\rm{H}}}{\bf{Dz}}$ controlling the sparsity of the estimation result and ${\left\| {{{\bf{Y}} - {\bf{W}}^H{\bf{{\hat H}X}}}} \right\|_{\rm F}}$ denoting the residue. In addition, $\lambda$ is the regularization parameter that controls the tradeoff between the sparsity and the data fitting error.

In the iterative reweighted method \cite{G12}, $\lambda$ is not fixed but updated in each iteration. To be specific, if the previous iteration is poorly-fitted, we will choose a smaller $\lambda$ to make the estimate sparser. On the other hand, if the previous iteration returns a well-fitted estimate and leads to a small residue, our method will choose a larger $\lambda$ to accelerate the searching for the best-fitting estimate. In the proposed algorithm, $\lambda$ is updated by
\vspace{-1mm}
\begin{equation}\label{equ:lambda}\
\vspace{-1mm}
\lambda={\rm min}\left(d/r^{(i)},\lambda_{\rm max}\right),
\end{equation}
where $d$ is a constant scaling factor, and $\lambda_{\rm max}$ is selected to make the problem well-conditioned, $r^{(i)}$ is the squared residue in the previous step, i.e.,
\vspace{-2mm}
\begin{equation}
\vspace{-1mm}
r^{(i)} = {\left\| {{\bf{Y}} - {\bf{W}}^H{{\bf{A}}_{\rm R}}\!\!\left(\! {\bm{\hat \theta} _{\rm R}^{(i)}}\! \right)\!{\rm{diag}}\!\!\left(\! {{{\bf{\hat z}}^{\!(i)}}} \!\right)\!{\bf{A}}_{\rm T}^{{H}}\!\!\left( \!{\bm{\hat \theta} _{\rm T}^{(i)}} \!\right)\!{\bf{X}}} \right\|^2_{\rm F}}.
\end{equation}
{The updation of $\lambda$ was discussed in [10] with more details.}

The proposed algorithm starts iteration at the angle domain grids. In the $i$-th iteration, our task is to search for new estimates ${\bm{\hat \theta}}_{\rm R}^{(i+1)}$ and ${\bm{\hat \theta}}_{\rm T}^{(i+1)}$ in the neighborhood of the previous estimates ${\bm{\hat \theta}}_{\rm R}^{(i)}$ and ${\bm{\hat \theta}}_{\rm T}^{(i)}$ to make the objective function ${S}^{(i)}$ become smaller. This searching can be accomplished via gradient descent method:
\vspace{-2mm}
\begin{equation}\label{equ:grad}
\begin{aligned}
    {\bm{\hat \theta}} _{\rm R}^{(i + 1)} = {\bm{\hat \theta}} _{\rm R}^{(i)} - \eta  \cdot \nabla _{{\bm{\theta}} _{\rm R}}{{S}_{\rm opt}^{(i)}}\left( {{\bm{\hat \theta}} _{\rm R}^{(i)}},{{\bm{\hat \theta}} _{\rm T}^{(i)}} \right),\\
    {\bm{\hat \theta}}_{\rm T}^{(i + 1)} = {\bm{\hat \theta}}_{\rm T}^{(i)} - \eta  \cdot \nabla _{{\bm{\theta}} _{\rm T}}{{S}_{\rm opt}^{(i)}}\left( {{\bm{\hat \theta}} _{\rm R}^{(i)}},{{\bm{\hat \theta}} _{\rm T}^{(i)}} \right),
\end{aligned}
\vspace{-2mm}
\end{equation}
where the gradients can be calculated according to Appendix B, and $\eta$ is the chosen step-length to make sure $S_{\rm opt}^{(i)}\!\!\left({\bm{\hat \theta}}_{\rm R}^{(i+1)}, {\bm{\hat \theta}}_{\rm T}^{(i+1)}\right)\! \le\! S_{\rm opt}^{(i)}\!\!\left({\bm{\hat \theta}}_{\rm R}^{(i)}, {\bm{\hat \theta}}_{\rm T}^{(i)}\right)$.
The estimates become more and more accurate during the iterative searching, until the new estimates are almost the same as the previous ones. With our proposed IR-based super-resolution channel estimation scheme, the estimates of $({\bm{\theta}}_{\rm R}, {\bm{\theta}}_{\rm T})$ can be moved from the initial on-grid coarse estimates to its actual off-grid positions, thus the super-resolution channel estimation can be realized.

It is worthy to point out that the sparsity level $L$ is unknown in practice. In the proposed scheme, the sparsity level can be initialized to be larger than the real channel sparsity. During the iteration process, the paths with too small path gains will be regarded as noise instead of real paths. Then, our algorithm prune these paths to make the result sparser. By iteratively pruning these paths, the estimated sparsity level will decrease to the real number of paths.

The computational complexity in each iteration lies in calculating the gradient in Step 5. The computational complexity to calculate the gradient is $\mathcal{O}\left(N_{\rm X}N_{\rm Y}(N_{\rm R}\!+\!N_{\rm T})L^2\right)$. As a result, the number of initial candidates $L^{(0)}$ is critical, and it should be as small as possible to make the computation affordable. The problem how to effectively select the initial ${\bm{\hat \theta}} _{\rm R}^{(0)}$ and ${\bm{\hat \theta}} _{\rm T}^{(0)}$ before the iteration will be discussed in the next section.
\vspace{-1.5mm}
\subsection{SVD-based Preconditioning}
In this section, we propose a singular value decomposition (SVD)-based preconditioning as shown in {\bf Algorithm \ref{alg:SVD}}, to reduce the computational complexity of the IR procedure in the proposed IR-based super-resolution channel estimation scheme. The proposed scheme can find the angle-domain grids nearest to the real AoAs/AoDs. Comparing to using all $N_{\rm R}N_{\rm T}$ angle-domain grids as initial candidates, the preconditioning can significantly reduce the computational complexity of the IR-based super-resolution channel estimation scheme.

Specifically, by applying SVD to the matrix $\bf Y$, we have ${{\bf Y}} = {\bf U\Sigma V}^{H}$, where ${\bf \Sigma} = {\rm diag}\left(\sigma _1, \sigma _2, \cdots, \sigma _{{\rm min}(N_{\rm X}, N_{\rm Y})} \right)\in {\mathbb{R}^{N_{\rm Y} \times N_{\rm X}}}$ whose diagonal entries $\sigma _1 \ge \sigma _2 \ge \cdots \ge \sigma _{{\rm min}(N_{\rm X}, N_{\rm Y})} \ge 0$ are the singular values of $\bf Y$, and ${\bf U}^H{\bf U}={\bf I}_{N_{\rm Y} \times N_{\rm Y}}$, ${\bf V}^H{\bf V}={\bf I}_{N_{\rm X} \times N_{\rm X}}$. From (\ref{equ:H1}) and (\ref{equ:Y}), we have
\vspace{-1.5mm}
\begin{equation}\label{equ:YSVD}
{{\bf Y}} = \left({\bf W}^{H}{{\bf{A}}_{\rm R}}\!\left( {{\bm{\theta} _{\rm R}}} \right) \right){\rm{diag}}\left( {\bf{z}} \right)\left({\bf{X}}^{H}{\bf A}_{\rm T}\!\left( {{\bm{\theta} _{\rm T}}} \right)\right)^{H}+{\bf N}.
\vspace{-1mm}
\end{equation}

As the noise is small, the largest $L$ singular values and their corresponding singular vectors are approximately determined by the $L$ paths, i.e., for $i=1,2,\cdots,L$, we have
\vspace{-1mm}
\begin{equation}\label{equ:UVpattern}
\begin{aligned}
&{\sigma}_i \approx \left|{z_{l_i}}\right|\left\|\!  {\bf W}^{\!H}\!{\bf{a}}_{\rm R}\!\!\left( {{\theta}^{\rm azi} _{{\rm R},l_i}}, \!{{\theta}^{\rm ele} _{{\rm R},l_i}}\right)\!\right\|_2\left\|\! {\bf X}^{\!H}\!{\bf{a}}_{\rm T}\!\!\left( {{\theta}^{\rm azi} _{{\rm T},l_i}},\!{{\theta}^{\rm ele} _{{\rm T},l_i}}\right)\!\right\|_2, \\
&{\bf u}_i \approx {\bf W}^{\!H}{\bf{a}}_{\rm R}\!\left( {{\theta}^{\rm azi} _{{\rm R},l_i}}, {{\theta}^{\rm ele} _{{\rm R},l_i}} \right)\!/\left\|  {\bf W}^{\!H}{\bf{a}}_{\rm R}\!\left( {{\theta}^{\rm azi} _{{\rm T},l_i}},{{\theta}^{\rm ele} _{{\rm T},l_i}} \right)\right\|_2, \\
&{\bf v}_i \approx {\bf X}^{H}{\bf{a}}_{\rm T}\left( {{\theta}^{\rm azi} _{{\rm R},l_i}}, {{\theta}^{\rm ele} _{{\rm R},l_i}}\right)/\left\|  {\bf X}^{H}{\bf{a}}_{\rm T}\left( {{\theta}^{\rm azi} _{{\rm T},l_i}},{{\theta}^{\rm ele} _{{\rm T},l_i}} \right)\right\|_2, \\
\end{aligned}
\vspace{-1mm}
\end{equation}
where ${\bf u}_i$ and ${\bf v}_i$ are the $i$-th column of $\bf U$ and $\bf V$, respectively, $\left\{l_1,l_2,\cdots,l_L\right\}$ is a permutation of $\left\{1,2,\cdots,L\right\}$.

Then, in steps 4-5 of {\bf Algorithm \ref{alg:SVD}}, we search for the coarse estimate of normalized AoAs (AoDs) in the finite set of angle-domain grids $\Omega_{\rm R}$ ($\Omega_{\rm T}$). {Take UPA as an example. For an $N_1\times N_2$ receiver array, the set of grids can be defined by $\Omega_{\rm R}=\left\{(i/N_1,j/N_2)|i=0,1,\cdots,N_1-1;j=0,1,\cdots,N_2-1\right\}$. We can similarly define $\Omega_{\rm T}$ for the transmitter.}

{In \cite{G12}, the initial candidates of {\bf Algorithm \ref{alg:Framwork}} are set to be all the grids, i.e., $L^{(0)}=N_{\rm R}N_{\rm T}$. The computational complexity is $\mathcal{O}\left(N_{\rm X}N_{\rm Y}(N_{\rm R}\!+\!N_{\rm T})N_{\rm R}^2N_{\rm T}^2\right)$, which is unaffordable when $N_{\rm R}$ and $N_{\rm T}$ are very large. Fortunately, with the proposed SVD-based preconditioning shown in {\bf Algorithm \ref{alg:SVD}}, the coarse estimates will be used as the initial candidates of {\bf Algorithm \ref{alg:Framwork}}, i.e., $L^{(0)}=N_{\rm init}\approx L$. Thus, the computational complexity after SVD preconditioning is $\mathcal{O}\left(N_{\rm X}N_{\rm Y}(N_{\rm R}\!+\!N_{\rm T})L^2\right)$, which is much lower than directly applying the scheme in \cite{G12}.}

\begin{algorithm}[!t]
{
\renewcommand{\algorithmicrequire}{\textbf{Input:}}
\renewcommand\algorithmicensure {\textbf{Output:} }
\caption{SVD-based preconditioning}
\label{alg:SVD}
\small
\begin{algorithmic}[1]
\REQUIRE
 Noisy received signals ${\bf Y}$, transmit pilot signals ${\bf X}$, combining matrix ${\bf W}$, and $N_{\rm init}$, the number of paths to detect.
\ENSURE
Coarse AoAs/AoDs estimates of the $N_{\rm init}$ paths.
\STATE {$\left[{\bf U}, {\bf \Sigma}, {\bf V}\right] = {\rm SVD}({\bf Y})$.}
\STATE {Take the first $N_{\rm init}$ columns, $\left\{{\bf u}_1,{\bf u}_2,\cdots,{\bf u}_{N_{\rm init}}\right\}$ from ${\bf U}$, and $\left\{{\bf v}_1,{\bf v}_2,\cdots,{\bf v}_{N_{\rm init}}\right\}$ from ${\bf V}$, which are correspondent to the $N_{\rm init}$ largest singular values.}
\FOR {$i = 1,2,\cdots,N_{\rm init}$}
    \STATE {$\left(\hat \theta _{{\rm R},i}^{{\rm azi}(0)},\hat \theta _{{\rm R},i}^{{\rm ele}(0)}\right)=\mathop{\arg\max}\limits_{\left(\theta _{{\rm R}}^{{\rm azi}},\theta _{{\rm R}}^{{\rm ele}}\right)\in \Omega_{\rm R}}{\bf u}_i^H{\bf W}^{\!H}{\bf{a}}_{\rm R}\!\left( {{\theta}^{\rm azi} _{{\rm R}}}, {{\theta}^{\rm ele} _{{\rm R}}} \right)$.}
    \STATE {$\left(\hat \theta _{{\rm T},i}^{{\rm azi}(0)},\hat \theta _{{\rm T},i}^{{\rm ele}(0)}\right)=\mathop{\arg\max}\limits_{\left(\theta _{{\rm T}}^{{\rm azi}},\theta _{{\rm T}}^{{\rm ele}}\right)\in \Omega_{\rm T}}{\bf v}_i^H{\bf X}^{\!H}{\bf{a}}_{\rm T}\!\left( {{\theta}^{\rm azi} _{{\rm T}}}, {{\theta}^{\rm ele} _{{\rm T}}} \right)$.}
\ENDFOR
\end{algorithmic}
}
\end{algorithm}

\vspace{-0.5mm}
\section{Simulation Results and Discussions}\label{S5}
\vspace{-0.5mm}
In this section, simulation results are provided to investigate the performance. We consider the mmWave massive MIMO system with hybrid precoding, where $L=3$, $d ={\lambda}/{2}$, $N_{\rm R}=N_{\rm T}=64$, $N_{\rm R}^{\rm RF}=N_{\rm T}^{\rm RF}=4$ and $N_{\rm X}=N_{\rm Y}=32$.
The path gains are assumed Gaussian, i.e., $\alpha_l\sim {\cal CN}(0 ,{\sigma_{\alpha} ^2})$. Each element of the transmitted pilots $\bf X$ satisfies $x_{i,j}={\sqrt { {\rho}/ {N_{\rm T}}}}e^{j\omega_{i,j}}$, where $\rho$ is the transmitted power, $\omega_{i,j}$ is the random phase uniformly distributed in $[0,2\pi)$. The signal-to-noise ratio (SNR) is defined by ${\rm SNR}=\frac{\rho\sigma_{\alpha}^2}{\sigma_{n}^2}$, where $\sigma_{n}^2$ is the noise variance. We consider the ULA geometry, so that the adaptive codebook-based channel estimation \cite{G7}, the auxiliary beam pair based channel estimation \cite{ABP}, and the OMP-based channel estimation \cite{G8} can be adopted for performance comparison.
\begin{figure}[tp!]
\begin{center}
\vspace*{-4mm}
\includegraphics[height=0.67\columnwidth, keepaspectratio]{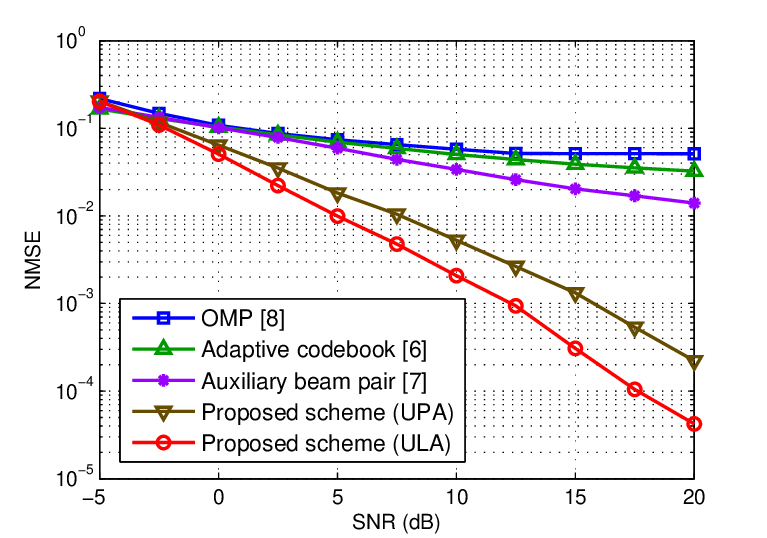}
\end{center}
\vspace*{-5mm}
\caption{NMSE performance comparison of different channel estimation schemes under NLoS channel.}
\label{fig:mse_vs_snr_nlos}
\vspace*{-4mm}
\end{figure}

\begin{figure}[tp!]
\begin{center}
\includegraphics[height=0.67\columnwidth, keepaspectratio]{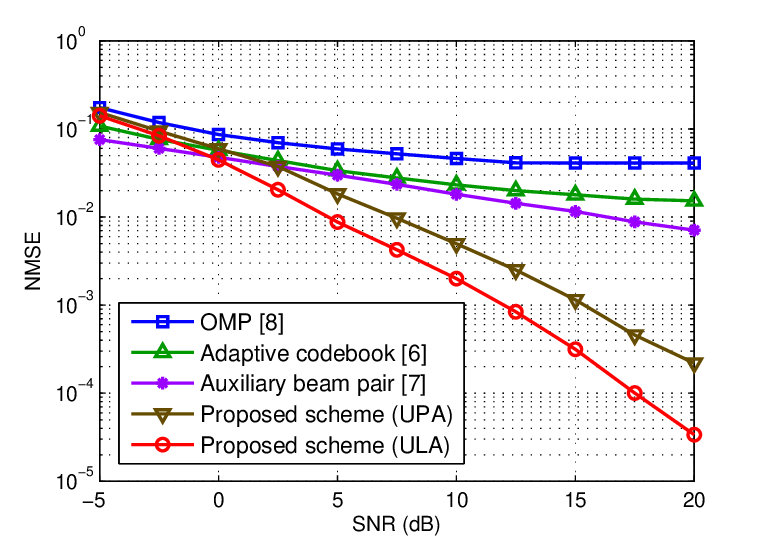}
\end{center}
\vspace*{-5mm}
\caption{NMSE performance comparison of different channel estimation schemes under LoS channel.}
\label{fig:mse_vs_snr_los}
\vspace*{-4mm}
\end{figure}

\begin{figure}[tp!]
\begin{center}
\includegraphics[height=0.67\columnwidth, keepaspectratio]{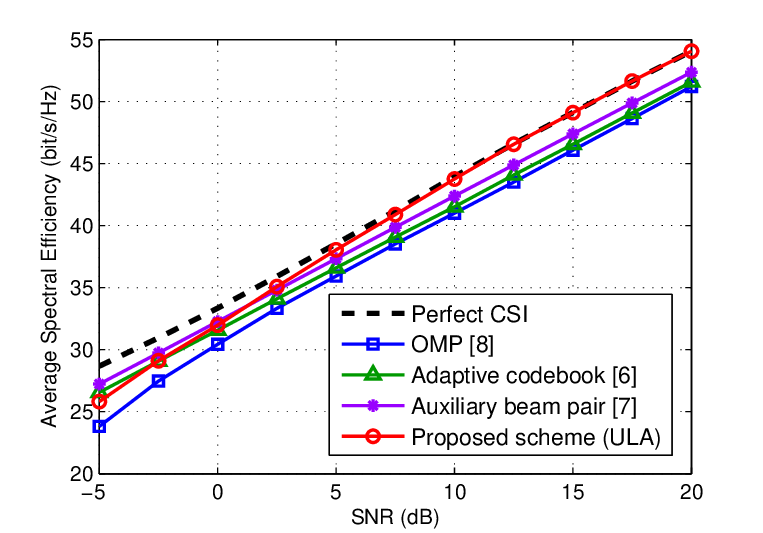}
\end{center}
\vspace*{-5mm}
\caption{Average spectral efficiency when different channel estimation schemes are used.}
\label{fig:mse_vs_T}
\vspace*{-5mm}
\end{figure}

{Fig.~\ref{fig:mse_vs_snr_nlos} and Fig.~\ref{fig:mse_vs_snr_los} compares the normalized mean square error (NMSE) performance against SNR, under none-line-of-sight (NLoS) and line-of-sight (LoS) channels, respectively. The Rician K-factor is $20$ dB in the LoS scenario. In both cases, the proposed scheme achieves much better NMSE performance when SNR becomes large. Moreover, we show the performance of the proposed scheme when UPA is considered. Both the transmitter and the receiver adopt 64-antenna UPA with 8 rows and 8 columns. We estimate the azimuth and elevation angles at both sides. We can observe that the proposed scheme is also able to achieve super-resolution channel estimation when UPA is used. Since the estimation errors of both azimuth and elevation angles contribute to the NMSE, under the same number of antennas and number of pilot overhead, the NMSE performance of UPA is higher than that of ULA.

Fig.~\ref{fig:mse_vs_T} compares the average spectral efficiency when different channel estimation schemes are used. {The spectral efficiency is evaluated in the hybrid precoding system \cite{G3}}. The case with ideal CSI was adopted as the upper bound for performance comparison. It can be observed that the proposed super-resolution channel estimation is able to approach this upper bound. This is because the angle resolution of the proposed scheme does not suffer from limited codebook size or angle quantization. Thus, we can conclude that the proposed scheme can achieve the super-resolution channel estimation.

{There is a tradeoff between the channel estimation accuracy and the computation complexity. The proposed scheme with SVD preconditioning is able to achieve much higher channel estimation accuracy, but has a higher computational complexity. The computational complexity of the proposed super-resolution channel estimation scheme with SVD precondistioning is $\mathcal{O}\left(N_{\rm X}N_{\rm Y}(N_{\rm R}\!+\!N_{\rm T})L^2\right)$. In comparison, the computational complexity of the OMP-based channel estimation \cite{G8} is $\mathcal{O}\left(N_{\rm X}N_{\rm Y}(N_{\rm R}\!+\!N_{\rm T})L\right)$. In order to achieve higher estimation accuracy, the increase in computational complexity is acceptable since $L$ is usually small for mmWave channels.}

\section{Conclusions}\label{S6}

In this paper, we have proposed an IR-based super-resolution channel estimation scheme for mmWave massive MIMO with hybrid precoding. Specifically, we have transformed the channel estimation problem to the optimization problem of a new objective function, which is the weighted summation of the sparsity and the data fitting error. The proposed scheme starts from the on-grid points in the angle domain, and iteratively moves them to the neighboring off-grid actual positions via gradient descent method. In addition, we have proposed an SVD-based preconditioning to reduce the computational complexity. Simulation results have confirmed that the proposed super-resolution channel estimation scheme can advance the state-of-art by estimating the off-grid AoAs/AoDs with much increased accuracy.
Angle estimation is the key of channel estimation for mmWave massive MIMO. Estimating the AoAs/AoDs with higher resolution is a practical way to realize higher spectral efficiency. For future work, it would be interesting to study other super-resolution channel estimation schemes with reduced complexity. In addition, super-resolution channel estimation under high mobility is an important yet challenging topic to be investigated.

\begin{appendices}

    \section{Optimization of $S$ in (\ref{equ:iteropt}) with regard to $\bf{z}$}\label{AppendixA}
    For notational conciseness, we ignore the superscript $(i)$ of $S^{(i)}$ and ${\bf D}^{(i)}$ in (\ref{equ:iteropt}), and use ${\bf A}_{\rm R}$, ${\bf A}_{\rm T}$ for ${\bf A}_{\rm R}\left({\bm \theta}_{\rm R}\right)$, ${\bf A}_{\rm T}\left({\bm \theta}_{\rm T}\right)$ respectively. Let ${{\bf{K}}_p} = {\bf{W}}^{\!H\!}{{\bf{A}}_r}{\rm{diag}}\left( {{\bf{A}}_t^{\rm{H}}{{\bf{x}}_p}} \right)$. In order to find the optimal $S\left( {{\bf{z}},{{\bm{\theta}} _{\rm R}},{{\bm{\theta}} _{\rm T}}} \right)$ with regard to ${\bf{z}}$, we can expand the objective function $S$ as
    \vspace{-2mm}
    \begin{small}
    \begin{equation}
    \begin{aligned}
    &S\left( {{\bf{z}},{{\bm{\theta}} _{\rm R}},{{\bm{\theta}} _{\rm T}}} \right) ={\lambda ^{{\rm{ - }}1}}{{\bf{z}}^{H}}{\bf{Dz}} +\sum\limits_{p = 1}^{N_{\rm X}}\left\| {{{\bf{y}}_p}-{\bf{W}}^{\!H\!\!}{{\bf{A}}_{\rm R}}{\rm{diag}}\left( {\bf{z}} \right){\bf{A}}_{\rm T}^{H}{{\bf{x}}_{p}}} \right\|^2_2\\
    &\!={\lambda ^{{\rm{ - }}1}}{{\bf{z}}^{H}}{\bf{Dz}} +\sum\limits_{p = 1}^{N_{\rm X}}\left( {{{\bf{y}}_p}-{\bf K}_p {\bf{z}}} \right)^H\!\left( {{{\bf{y}}_p}-{\bf K}_p {\bf{z}}} \right)\\
    &\!={{\bf{z}}^{\!H\!\!}}\!\left( \!{{\lambda ^{ \!-\!\!1}}{\bf{\!D}} \!+\!\! \sum\limits_{p = 1}^{N_{\rm X}} \!{{\bf{K\!}}_p^{H}\!{{\bf{K\!}}_p}} } \! \right)\!{\bf{z}}\!-\!{{\bf{z}}^{\!H\!\!\!}}\left(\sum\limits_{p = 1}^{N_{\rm X}}\!{{\bf{K\!}}_p^{H}\!{{\bf{y\!}}_{p}}}\! \right)\!\!-\!\!\left(\sum\limits_{p = 1}^{N_{\rm X}}\! {{\bf{y\!}}_p^{H}\!{{\bf{K\!}}_{p}}}\!\! \right)\!{{\bf{z}}}\!+\!\!\sum\limits_{p = 1}^{N_{\rm X}}\!{{\bf{y\!}}_{p}^{H}\!{{\bf{y\!}}_{p}}},\\
    \end{aligned}
    \end{equation}
    \end{small}
    Then, we can obtain the partial derivative by
    \vspace{-2mm}
    \begin{small}
    \begin{equation}
    \begin{aligned}
    \frac{{\partial S\left( {{\bf{z}},{{\bm{\theta}} _{\rm R}},{{\bm{\theta}} _{\rm T}}} \right)}}{{\partial {\bf{z}}}} = &~{{\bf{z}}^{H}}\!\!\left( {{\lambda ^{ - 1}}{\bf{D}} + \sum\limits_{p = 1}^{N_{\rm X}} {{\bf{K}}_p^{H}{{\bf{K}}_p}} } \right) - \left( {\sum\limits_{p = 1}^{N_{\rm X}} {{\bf{y}}_p^{H}{{\bf{K}}_p}} } \right).
    \end{aligned}
    \vspace{-0.5mm}
    \end{equation}
    \end{small}
    By setting the derivative to zero, the minimum point ${\bf{z}}$ and the corresponding minimum value of $S\left( {{\bf{z}},{{\bm{\theta}} _{\rm R}},{{\bm{\theta}} _{\rm T}}} \right)$ as the function of ${{\bm{\theta}} _{\rm R}}$ and ${{\bm{\theta}} _{\rm T}}$ can be obtained as
    \vspace{-2mm}
    \begin{small}
    \begin{equation}
        {{\bf{z}}_{\rm opt}}\!\left( {{{\bm{\theta}} _{\rm R}},{{\bm{\theta}} _{\rm T}}} \right) = {\left( {{\lambda ^{ \!- 1}}{{\bf{D}}}\! +\! \sum\limits_{p = 1}^{N_{\rm X}} {{\bf{K}}_p^{H}{{\bf{K}}_p}} } \right)^{ \!\!\!- 1}}\!\!\left( {\sum\limits_{p = 1}^{N_{\rm X}} {{\bf{K}}_p^{H}{{\bf{y}}_p}} } \right),
    \end{equation}
    \vspace{-2mm}
    \begin{equation}
    \begin{aligned}
        {{S}_{\rm opt}}\!\left( {{{\bm{\theta}} _{\rm R}},{{\bm{\theta}} _{\rm T}}} \right) =  - \!\left( {\sum\limits_{p = 1}^{N_{\rm X}} {{\bf{K}}_p^{H}{{\bf{y}}_p}} } \right)^{\!H} \!\!\! {\left( {{\lambda ^{\! - 1}}{{\bf{D}}}\! + \!\sum\limits_{p = 1}^{N_{\rm X}} {{\bf{K}}_p^{H}{{\bf{K}}_p}} } \right)^{ \!\!\!- \!1}}\\
        \!\cdot\!\left( {\sum\limits_{p = 1}^{N_{\rm X}} {{\bf{K}}_p^{H}{{\bf{y}}_p}} } \right) + \sum\limits_{p = 1}^{N_{\rm X}} {{\bf{y}}_{p}^{H}{{\bf{y}}_{p}}}.
    \end{aligned}
    \end{equation}
    \end{small}

    \vspace{-2mm}
    \section{Gradient of $S_{\rm opt}\left({{{\bm{\theta}} _{\rm R}},{{\bm{\theta}} _{\rm T}}} \right)$}
    Denote ${\bf v}\!=\!{\sum_{p = 1}^{N_{\rm X}} {{\bf{K}}_p^{H}{{\bf{y}}_p}} }$, ${\bf A}\!=\!{\lambda ^{ \!-\! 1}}{\bf{D}}+{\sum_{p = 1}^{N_{\rm X}} {{\bf{K}}_p^{H}{{\bf{K}}_p}} }$, we have {$S_{\rm opt}\!=\!-{\bf v}^H{\bf A}^{\!-1}{\bf v}+{\sum_{p = 1}^{N_{\rm X}} {{\bf{y}}_p^{H}{{\bf{y}}_p}}}$}. Take partial derivative with respect to ${\theta _{{\rm R},l}}$, we have
    \vspace{-0.5mm}
    \begin{small}
    \begin{equation}
    \begin{aligned}
    \frac{{\partial {S_{\rm opt}}}}{{\partial {\theta _{{\rm R},l}}}} &=  - \frac{{\partial {{\bf{v}}^{H}}}}{{\partial {\theta _{{\rm R},l}}}}{{\bf{A}}^{ \!-\! 1}}{\bf{v}} \!-\! {{\bf{v}}^{H}}\frac{{\partial {{\bf{A}}^{ \!-\! 1}}}}{{\partial {\theta _{{\rm R},l}}}}{\bf{v}} \!-\! {{\bf{v}}^{H}}{{\bf{A}}^{ \!-\! 1}}\frac{{\partial {\bf{v}}}}{{\partial {\theta _{{\rm R},l}}}}\\
    &=- \frac{{\partial {{\bf{v}}^{H}}}}{{\partial {\theta _{{\rm R},l}}}}\!{{\bf{A}}^{ \!-\! 1}}{\bf{v}} \!+\! {{\bf{v}}^{H}}\!{{\bf{A}}^{ \!-\! 1}}\!\frac{{\partial {\bf{A}}}}{{\partial {\theta _{{\rm R},l}}}}{{\bf{A}}^{ \!-\! 1}}{\bf{v}} \!-\! {{\bf{v}}^{H}}\!{{\bf{A}}^{ \!-\! 1}}\!\frac{{\partial {\bf{v}}}}{{\partial {\theta _{{\rm R},l}}}},
    \end{aligned}
    \vspace{-1.5mm}
    \end{equation}
    \end{small}
    \vspace{-1mm}
    where
    \begin{small}
    \begin{equation}
    \begin{aligned}
    \frac{{\partial {\bf{v}}}}{{\partial {\theta _{{\rm R},l}}}} &\!=\! \sum\limits_{p = 1}^{N_{\rm X}} \!{\frac{{\partial {\bf{K\!}}_p^H}}{{\partial {\theta _{{\rm R},l}}}}\! {{\bf{y}}_p}},~
    \frac{{\partial {\bf{A}}}}{{\partial {\theta _{{\rm R},l}}}} \!=\! \sum\limits_{p = 1}^{N_{\rm X}} {\left( {\frac{{\partial {\bf{K\!}}_p^{H}}}{{\partial {\theta _{{\rm R},l}}}}\! {{\bf{K}}_p} \!+\! {\bf{K\!}}_p^{H} \! \frac{{\partial {{\bf{K}}_p}}}{{\partial {\theta _{{\rm R},l}}}}} \right)},\\
    \frac{{\partial {{\bf{K}}_p}}}{{\partial {\theta _{{\rm R},l}}}} &\!=\! \left[ {\begin{array}{*{20}{c}}
   {\!\!\bf 0} & {\!\!\!\cdots\!\!\!} & {\bf 0\!\!} & {{\bf{W}}^H\frac{{\partial {{\bf{a}}_{\rm R}}\left( {{\theta _{{\rm R},l}}} \right)}}{{\partial {\theta _{{\rm R},l}}}}{\bf{a}}_{\rm T}^{H}\!\!\left( {{\theta _{{\rm T},l}}} \right){{\bf{x}}_p}} & {\!\!\bf 0} & {\!\!\!\cdots\!\!\!} & {\bf 0\!\!}  \\
   \end{array}} \right].
    \end{aligned}
    \end{equation}
    \end{small}
\end{appendices}

\end{document}